\documentclass{mn2e}
\usepackage{color}
\usepackage{graphicx}
\usepackage{dcolumn}
\usepackage{bm}
\usepackage{lscape}
\usepackage{epsfig}
\input{psfig}

\title[Extending the $M_{\rm bh}$--$\sigma$ diagram]
{Extending the $M_{\rm bh}$--$\sigma$ diagram with dense nuclear star clusters}

\author[Alister W.\ Graham]
{Alister W.\ Graham$^1$\thanks{AGraham@astro.swin.edu.au}, \\
$^1$ Centre for Astrophysics and Supercomputing, Swinburne University
of Technology, Hawthorn, Victoria 3122, Australia.
}

\date{Submitted 08 Aug, 2011} 
\begin{document}
\label{firstpage}
\maketitle

\begin{abstract}

Four new nuclear star cluster masses, $M_{\rm nc}$, plus seven upper limits, are provided for
galaxies with previously determined black hole masses, $M_{\rm bh}$.
Together with a sample of 64 galaxies with direct $M_{\rm bh}$  
measurements, 13 of which additionally now have $M_{\rm nc}$ 
measurements rather than only upper limits, plus an additional 29 dwarf galaxies
with available $M_{\rm nc}$
measurements and velocity dispersions $\sigma$, an
($M_{\rm bh} + M_{\rm nc}$)--$\sigma$ diagram is constructed.
Given that major dry galaxy merger events preserve the $M_{\rm bh}/L$ ratio,
and given that $L \propto \sigma^5$ for luminous galaxies, it is first noted 
that the observation $M_{\rm bh} \propto \sigma^5$ is 
consistent with expectations.
For the fainter elliptical galaxies it is known that $L \propto \sigma^2$, and
assuming a constant $M_{\rm nc}/L$ ratio (Ferrarese et al.), 
the expectation that $M_{\rm nc} \propto \sigma^2$ 
is in broad agreement with our new observational result that $M_{\rm nc} \propto
\sigma^{1.57\pm0.24}$.
This exponent is however in contrast to the value of $\sim$4 which
has been reported previously and interpreted in terms of a regulating feedback
mechanism from stellar winds. 

Finally, it is predicted that host galaxies fainter than $M_B \sim -20.5$ mag
(i.e.\ those not formed in dry merger events) which follow the relation
$M_{\rm bh} \propto \sigma^5$, and are thus not `pseudobulges', should 
not have a constant $M_{\rm bh}/M_{\rm host}$
ratio but instead have $M_{\rm bh} \propto L^{5/2}_{\rm host}$.
It is argued that the previous near linear $M_{\rm bh}$--$L$ and $M_{\rm
bh}$--$M_{\rm spheroid}$ relations have been biased by the sample
selection of luminous galaxies, and as such should not be used to constrain
the co-evolution of supermassive black holes in galaxies other than those
luminous few built by major dry merger events.

\end{abstract}


\section{Introduction}

Supermassive black holes have long been known to exist at the centers of large
galaxies (e.g.\ Lynden-Bell 1969; Wolfe \& Burbidge 1970; Sargent et al.\
1978).  Intriguingly, scaling relations between their masses and several
global properties of the host spheroid (e.g.\ Kormendy \& Richstone 1995;
Magorrian et al.\ 1998; Ferrarese \& Merritt 2000; Gebhardt et al.\ 2000;
Graham et al.\ 2001; Marconi \& Hunt 2003) may be the result of feedback
mechanisms in which the central black hole regulates the growth of the
surrounding bulge, rather than vice-versa (e.g.\ Silk \& Rees 1998; Haehnelt,
Natarajan \& Rees 1998; de Lucia et al.\ 2006; Antonuccio-Delogu \& Silk
2010).

In early-type dwarf galaxies and the bulges of late-type galaxies, dense
nuclear star clusters appear to dominate at the expense of massive black holes
(Valluri et al.\ 2005; Ferrarese et al.\ 2006a; Wehner \& Harris 2006).  The
existence of scaling relations between the luminosity and stellar mass of
these star clusters and their host spheroid (e.g.\ Graham \& Guzm\'an 2003;
Balcells et al.\ 2003, 2007; Grant et al.\ 2005) similarly suggests that a
physical mechanism may be controlling their growth, possibly based on some
regulating feedback process (e.g.\ King 2005; McLaughlin et al.\ 2006;
Hueyotl-Zahuantitla et al.\ 2010).  Or perhaps instead some other activity prevails,
such as cluster inspiral (e.g.\ Tremaine, Ostriker \& Spitzer 1975; Bekki 2010, 
Agarwal \& Milosavljevi{\'c} 2011), possibly coupled with gas dissipation and new star formation 
(Hartmann et al.\ 2011). 
% Agarwal \& Milosavljevi{\'c} 2011).  

Coupled with the above observational relations
% , some of which are still under development, 
is the observation that many nuclear star clusters in
intermediate-mass spheroids (of stellar mass $10^{8} < M_{\rm sph,*}/M_{\odot}
< 10^{10}$) harbour massive black holes themselves (e.g.\ Graham \& Driver
2007; Gonz{\'a}lez Delgado et al.\ 2008; Seth et al.\ 2008, 2010; Gallo et
al.\ 2010; Neumayer \& Walcher 2012).  An attempt to quantify the coexistence of these two types of
galactic nuclei was provided by Graham \& Spitler (2009) who revealed how
their (i) mass ratio and (ii) combined mass relative to their host spheroid's
stellar mass, changed as a function of host spheroid stellar mass.
Such dual nuclei are exciting for a number of reasons, including UV/X-ray
flaring events as infalling stars are tidally disrupted by the black hole
(e.g.\ Komossa \& Merritt 2008; Lodato et al.\ 2008; Rosswog et al.\ 2008;
Maksym, Ulmer \& Eracleous 2010) and the increased expectation for the
discovery of gravitational radiation as stellar mass black holes and neutron
stars inspiral toward the central supermassive black hole of these dense,
compact star clusters (Mapelli et al.\ 2011).

If nucleated galaxies, i.e.\ those with nuclear star clusters, were
participants in an hierarchical universe (White \& Frenk 1991), then their
dense nuclei must have eventually been replaced by massive black holes as
they, the host galaxies, grew into massive elliptical galaxies.  Bekki \&
Graham (2010) have argued that the gravitational scouring which ensues from a
coalescing binary supermassive black hole after a galaxy merger event
(Begelman, Blandford \& Rees 1980; Ebisuzaki, Makino \& Okumura 1991; Graham
2004; Merritt, Mikkola \& Szell 2007), must first be preceded by the
destruction of these nuclear star clusters.  They have revealed that binary
supermassive black holes can effectively `heat' the newly-merged star
clusters, causing them to eventually evaporate into the host spheroid.  Such a
scenario suggests a connection-of-sorts between nuclear star clusters and
massive black holes in intermediate mass spheroids.  Other, perhaps yet
unthought of, processes may also be operating.
This Letter explores potential connections by expanding upon the association
between black hole mass and host galaxy velocity dispersion, the $M_{\rm
bh}$--$\sigma$ diagram (Ferrarese \& Merritt 2000; Gebhardt et al.\ 2000;
Graham et al.\ 2011), by including nuclear star clusters.

In section~\ref{Sec_light} we provide some insight into the expected relations
in the ($M_{\rm bh} + M_{\rm nc}$)--$\sigma$ diagram via reference to the
galaxy luminosity-(velocity dispersion) relation for dwarf and ordinary
elliptical galaxies (Davies et al.\ 1983) and the galaxy-(nuclear star
cluster) luminosity relation for spheroids (Graham \& Guzm\'an 2003; Balcells
et al.\ 2007).
We also build on the ($M_{\rm bh} + M_{\rm nc}$)--$\sigma$ diagram from Graham
et al.\ (2011) by identifying and including new galaxies that host both a
nuclear star cluster and a supermassive black hole (Section~\ref{Sec_data}).
We additionally include those galaxies from Ferrarese et al.\ (2006a) with
nuclear star cluster masses that populate the low-mass end of the diagram.
In Section~\ref{Sec-R_D} we present our findings, notably that the expected relation
$M_{\rm nc} \propto \sigma^2$ appears consistent with the data.  This exponent
of 2 is dramatically different to the value of $4.27\pm0.61$ advocated
previously (Ferrarese et al.\ 2006a), and suggests that theories developed to
match the previous relation may need reconsideration. 
Section~\ref{Sec_Predict} goes on to present an exciting and significantly new 
prediction for the $M_{\rm bh}$--$M_{\rm sph}$ and $M_{\rm bh}$--luminosity relations 
for spheroids 
fainter than $M_B \sim -20.5$ mag, i.e.\ those
% without partially-depleted cores 
thought to have not formed from major, dissipationless, galaxy merger events
(e.g.\ Davies et al.\ 1983; Faber et al.\ 1997).

\section{Expectations}\label{Sec_light}

% $L \propto \sigma^4$ (Es), $\sigma^5$ luminous Es (ref), and $\sigma^2$ (dEs).\\
% Es: $M_{\rm bh} \propto L$ (dry merging), thus $L \propto \sigma^5$\\ 
% dEs: $M_{\rm nc} \propto L^{0.7}, L^{0.7} \propto \sigma^{2\times0.7} = \sigma^{1.4}$.

From pre-existing scaling relations it is possible 
to predict the slope of the relation between nuclear cluster mass and host
spheroid velocity dispersion: the $M_{\rm nc}$--$\sigma$ relation.  It
is also possible to predict a slope for the $M_{\rm bh}$--$\sigma$
relation at the high-mass end where nuclear clusters do not exist and dry
galaxy merging is thought to occur. 
% 
%\vspace{2mm}

The luminosity $L$ of dwarf elliptical galaxies 
(or more broadly elliptical galaxies without depleted cores) 
is such that $L\propto \sigma^2$ (Davies et al.\ 1983; Held et al.\
1992), while for big elliptical galaxies (with $\sigma \ga 200$ km s$^{-1}$)
the exponent is known to have a value of 5 (Schechter 1980; Malumuth \&
Kirshner 1981).  When including samples of intermediate-mass elliptical
galaxies (with $100 \la \sigma <$ 170-200 km s$^{-1}$) with the big
elliptical galaxies, the average
exponent has the more commonly known value of 3 to 4 (Faber \& Jackson 1976;
Tonry 1981).
Following Davies et al.'s (1983) identification of the transition in the
$L$--$\sigma$ relation at $M_B \approx -20.5$ B-mag ($\sigma \approx 200$ km
s$^{-1}$), where they noted that a number of other physical properties changed
behavior, Matkovi\'c \& Guzm\'an (2005, see also de Rijcke et al.\ 2005)
connected this transition with the onset of dry galaxy merging in the brighter
galaxies.

Provided there are no significant gravitational 
ejections of supermassive black holes from massive galaxies
(e.g.\ Gualandris \& Merritt 2008), then at the high-mass end where
dry galaxy merging is thought to occur --- 
involving galaxies with equal 
$M_{\rm bh}/M_{\rm sph}$ ratios (H\"aring \& Rix 2004) --- 
the combined supermassive black hole mass and the merged host galaxy 
luminosity and mass, must increase in lock step.  That is, the slope of the
$M_{\rm bh}$--$L$ relation must be equal to 1, as is 
observed for samples dominated by luminous galaxies (Marconi \& Hunt 2003; Graham 2007).
Consequently, the slope of the $L$--$\sigma$ relation for galaxies built by
such dry merging (with $M_B \la -20.5$ B-mag and $\sigma \ga$ 200 km 
s$^{-1}$) will therefore equal the slope of the $M_{\rm bh}$--$\sigma$ relation
over this same mass range.  Given that $L \propto \sigma^5$, one has (the
prediction) that $M_{\rm bh} \propto \sigma^5$, which is what is observed for
massive ``core'' galaxies (Hu 2008; Graham et al.\ 2011; see also 
Ferrarese \& Merritt 2000 and Merritt \& Ferrarese 2001). 

At the low-mass end, Graham \& Guzm\'an (2003) have
revealed that the nuclear cluster luminosity, 
and in turn stellar mass, 
$M_{\rm nc}$, in dwarf elliptical galaxies scales with the galaxy luminosity
$L$ such that $M_{\rm nc} \propto L^{0.87\pm0.26}$.  Given that $L \propto
\sigma^2$ in dwarf elliptical galaxies, one has that $M_{\rm nc} \propto
\sigma^{1.74\pm0.52}$, or, {\it roughly} that $M_{\rm nc} \propto \sigma^{2}$.
% 
% With a larger sample, including bulges from disc galaxies, Balcells et al.\
% (2007) reported the relation $M_{\rm nc} \propto L^{0.76\pm0.13}$,
% suggesting the $M_{\rm nc}$--$\sigma$ relation may have an exponent less
% than 2. 

Another way to predict the outcome is to note that if the ratio
of $(M_{\rm bh} + M_{\rm nc})$ to host spheroid luminosity $L$ is
constant (Ferrarese et al.\ 2006a), then the bent $L$--$\sigma$ relation 
(Davies et al.\ 1983) maps directly into a 
bent ($M+M$)-$\sigma$ relation, with slopes of 2 and 5 at the low- and
high-mass end respectively.  We note that this bent $(M+M)$--$\sigma$
relation has been predicted before 
% 
% (Graham\footnote{The penultimate sentence of Graham (2007b) contains a typo 
% and should have read $L \propto M_{\rm bh}^{0.5}$.} 2007b, his Appendix~A; 
% Bernardi et al.\ 2007;
% 
(e.g.\ Graham \& Driver 2007, their
section 3.2; Graham 2008b, their section~2.2.2) but curiously is at 
odds with Ferrarese et al.\ (2006a) who reported a slope of 
% 4.27$\pm$0.61 
$\sim$4 for the $M_{\rm nc}$--$\sigma$ relation.

\section{Data}\label{Sec_data}

\begin{table*}
\caption{Extension of Graham \& Spitler's (2009) Table~1 for galaxies
  with a direct supermassive black hole mass measurement (from the
  compilation by Graham 2008b and Graham et al.\ 2011) {\it and} a
  nuclear star cluster.  All galaxies that are likely to have both a
  supermassive black hole and a nuclear cluster, based upon their 
  'goldilocks' host spheroid stellar 
 mass (see Graham \& Spitler 2009) are included.}
\label{Tab1}
\begin{tabular}{@{}llccll@{}}
\hline
Galaxy    & Type & Dist. &    $M_{\rm bh}$         & Mag$_{\rm nc}$                           & $ M_{\rm nc}$     \\
          &      & Mpc  &  $10^7 [M_{\odot}]$      &     mag                                  & $10^7 [M_{\odot}]$  \\
\hline
NGC~1300  & SBbc & 20.7 &  $7.3^{+6.9}_{-3.5}$     &    ...                                   &  8.7$^A$ \\
NGC~2549  & SB0  & 12.3 &  $1.4^{+0.2}_{-1.3}$     & $m_{F702W} = 17.6$$^B$                   &  1.1  \\
NGC~3585  & S0   & 19.5 &  $31^{+14}_{-6}$         & $m_{F555W} = 20.5$$^C$                   &  0.4  \\  
NGC~4026  & S0   & 13.2 &  $18^{+6}_{-3}$          & $m_{F555W} = 18.4$$^C$                   &  1.3  \\  
\multicolumn{6}{c}{Upper limits on nuclear star cluster mass} \\
NGC~1316  & SB0  & 18.6 &  $15.0^{+7.5}_{-8.0}$    & $m_V > 19.9$$^D$                         &  $< 0.8$ \\
NGC~2787  & SB0  &  7.3 &  $4.0^{+0.4}_{-0.5}$     & $m_{F555W} > 17.25^{+0.17}_{-0.10}$$^E$  &  $< 1.5$ \\
NGC~3227  & SB   & 20.3 &  $1.4^{+1.0}_{-0.6}$     & $m_H > 15.7\pm0.2^F$                     &  $< 2.2$ \\
NGC~3245  & S0   & 20.3 &  $20^{+5}_{-5}$          & $m_{F547M} > 17.61^{+0.15}_{-0.11}$$^E$  &  $< 8.4$ \\
NGC~3489  & SB0  & 11.7 &  $0.58^{+0.08}_{-0.08}$  & $m_H > 12.7$$^G$                         &  $< 13$  \\
NGC~4459  & S0   & 15.7 &  $6.8^{+1.3}_{-1.3}$     & $m_{F555W} > 17.40^{+0.24}_{-0.14}$$^E$  &  $< 5.8$ \\
NGC~4596  & SB0  & 17.0 &  $7.9^{+3.8}_{-3.3}$     & $m_{F606W} > 17.97^{+0.14}_{-0.08}$$^E$  &  $< 4.0$ \\
\multicolumn{6}{c}{Unknown nuclear star cluster mass} \\
Circinus  & Sb   &  2.8 &  $0.11^{+0.02}_{-0.02}$  & \multicolumn{2}{l}{unknown, dusty Sy2 nucleus$^H$} \\
IC~2560   & SBb  & 40.7 &  $0.44^{+0.44}_{-0.22}$  & \multicolumn{2}{l}{unknown, dusty Sy2 nucleus$^I$} \\
NGC~224   & Sb   & 0.74 &  $14^{+9}_{-3}$          & \multicolumn{2}{l}{two nuclear discs$^J$} \\  
NGC~1068  & Sb   & 15.2 &  $0.84^{+0.03}_{-0.03}$  & \multicolumn{2}{l}{unknown, Sy2 nucleus$^K$} \\
NGC~3079  & SBcd & 20.7 &  $0.24^{+0.24}_{-0.12}$  & \multicolumn{2}{l}{unknown, dusty Sy2 nucleus$^L$} \\
NGC~3393  & SBab & 55.2 &  $3.4^{+0.2}_{-0.2}$     & \multicolumn{2}{l}{unknown, dusty Sy2 nucleus$^M$} \\ 
NGC~3998  & S0   & 13.7 &  $22^{+19}_{-16}$        & \multicolumn{2}{l}{unknown, AGN dominates$^E$} \\
NGC~4258  & SBbc &  7.2 &  $3.9^{+0.1}_{-0.1}$     & \multicolumn{2}{l}{unknown, Sy2 AGN dominates$^N$} \\
NGC~4261  & E2   & 30.8 &  $52^{+10}_{-11}$        & \multicolumn{2}{l}{unknown, Sy3 AGN dominates$^E$} \\
NGC~4486a & E2   & 17.0 &  $1.3^{+0.8}_{-0.8}$     & \multicolumn{2}{l}{nuclear stellar disc$^O$} \\
NGC~4945  & SBcd &  3.8 &  $0.14^{+0.14}_{-0.07}$  & \multicolumn{2}{l}{Sy2 $+$ dusty nuclear starburst$^P$} \\
NGC~5128  & S0   &  3.8 &  $4.5^{+1.7}_{-1.0}$     & \multicolumn{2}{l}{unknown, Sy2 AGN dominates$^Q$} \\
NGC~7582  & SBab & 22.0 &  $5.5^{+2.6}_{-1.9}$     & \multicolumn{2}{l}{unknown, Sy AGN dominates$^R$} \\
%
% Cygnus A  Sy2
% NGC 4374  Sy2
% NGC~4486  Sy/S3
% NGC~6251  Sy2
% NGC~2974  Sy2  
% NGC~4552  Sy2
\hline
\end{tabular}

\noindent
References:
$^A$ Atkinson et al.\ (2005, their Table~2, integrating their inner component to 10$r_b \approx 1\arcsec$); 
$^B$ From our NC$+$S\'ersic$+$exponential analysis of the light profile in Rest et al.\ (2001), using $M/L_{F702W}=1.5$; 
$^C$ From our NC$+$S\'ersic$+$exponential analysis of the light-profile in Lauer et al.\ (2005), using $M/L_{F555W}=2.0$; 
$^D$ Lauer et al.\ (2005), NGC~1316 = Fornax A, AGN contamination, $M/L_V=2.5$ used here;
$^E$ Gonzalez-Delgado et al.\ (2008), $M/L=2.5$ used here, nuclear cluster masses are upper limits due to AGN contamination; 
$^F$ Carollo et al.\ (2002), may have starburst plus Sy1.5 AGN contamination, $M/L_H=0.5$ used here; 
$^G$ From our NC$+$S\'ersic$+$exponential analysis of this Sy2
galaxy's light-profile in Nowak et al.\ (2010; their Figure~9), using $M/L_H=0.56$;
% upper limit because there is an AGN in N3489 which may contribute flux. 
$^H$ Prieto et al.\ 2004, Mu{\~n}oz-Mar{\'{\i}}n et al.\ (2007), Tristram et al.\ 2007;
$^I$ Peng et al.\ (2006), Mu{\~n}oz-Mar{\'{\i}}n et al.\ (2007); 
$^J$ Peterson (1978); 
$^K$ Davies et al.\ (2007, their Fig.22) uncalibrated light profile reveals a nuclear point source within 
0.1-0.2 arcseconds, atop of the 1 arcsecond (70 pc) nuclear disc in NGC~1068.
$^L$ Cecil et al.\ (2001);
$^M$ Cooke et al.\ (2000);
$^N$ Pastorini et al.\ (2007); % Siopis et al.\ (2009); 
$^O$ Kormendy et al.\ (2005), Ferrarese et al.\ (2006b: NGC~4486a = VCC~1327), Prugniel et al.\ (2011); 
$^P$ Marconi et al.\ (2000);
$^Q$ Radomski et al.\ (2008); 
% see also H\"aring-Neumayer et al. 2006, Meisenheimer et al. 2007
$^R$ Bianchi et al.\ (2007), Wold \& Galliano (2006); 
light-profile given by Rest et al.\ (2001), $M/L_{F702W}=1.5$ used here;
\end{table*}

% NGC~3227: $m_V$=17.3(0.3), $m_j$=16.0(0.2), $m_H$=15.7(0.2), Carollo et al.\ (2002).

% While Atkinson et al.\ (2005) provide a two-component fit to the central
% region of the inactive galaxy NGC~1300, their double power-law description of
% the inner component --- the nuclear star cluster --- is questionable due to
% the way their model eats into the bulge at large radii: integrating their
% fitted density model to three arcseconds gives an unrealistically high mass of
% $2\times10^8 M_{\odot}$ for the nuclear component.

% 6 dual SMBH-NC nuclei in Virgo: Gallo et al. (2010, ApJ 714 25).
% % NGC 404: ? Seth et al. (2010, ApJ, 714 713)

% For NGC 2787, Peng et al.\ (2002) gave m_{NC, F547M} > 19.8 mag (bad Nuker fits). 

% NGC 253 has lots of circumnuclear star clusters but no nuclear star cluster 
% (Fer/'andez-Ontiveros et al.\ 2008 AND arXiv:1005.1645) 

% NGC 524 may be a core galaxy (Byun et al. 1996, Quillen et al. 2001)
% NGC 2549: poorly fit light profile in Rest et al. (2001).

% NGC 5576 appears not to be nucleated, a fit to the light profile gives
% n=3.4 right into the centre. 

% NGC~1068 has a parsec scale disc/torus (Weigelt et al.\ 2004; Jaffe et
%     al. 2004; Davies et al.\ 2007 - with possible NC)

% Circinus has a parsec scale disc (Prieto et al.\ 2004; Tristram et al. 2007)

% Omega-Cen: Miocchi 2010, A\&A, 514, Article.52
% van der Marel on center issue.   Gebhardt et al. has a rebuttal.

% xxx maybe some data points in arXiv:1112.1417

The black hole masses for 64 galaxies have been taken from Graham
(2008b, his table~1) and Graham et al.\ (2011, their table~1).  The
velocity dispersions have also been obtained from the tables in these papers, with
the exception that this Letter uses a host spheroid velocity
dispersion of 55 km s$^{-1}$ for M32 (Chilingarian 2011, in prep.).
The previously tabulated central velocity dispersion of 72 km s$^{-1}$
for this nearby galaxy is elevated by the stellar dynamics close to
the spatially well-resolved black hole.
% (75 used by G\"ultekin et al. 2009). 

\begin{figure}
\includegraphics[angle=270,scale=0.36]{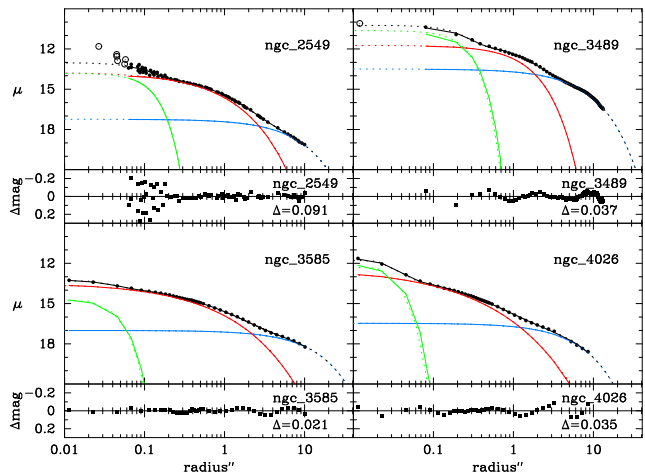}
\caption{
The magnitude of the nuclear star cluster is measured 
relative to the inward extrapolation of the available outer galaxy light 
distribution --- which has been modelled as the sum of two components: a
S\'ersic bulge plus an exponential disc.  Residual profiles, and the root mean
square (rms) scatter $\Delta$, are shown in the lower panels. 
The light profile data have come from the sources listed in Table~\ref{Tab1}. 
}
\label{Fig1}
\end{figure}

As noted by Graham \& Spitler (2009), many of these galaxies also house
nuclear star clusters.  
In the linear regression which follows, we do however exclude 
NGC~4564 (whose nuclear star cluster mass is not yet available) and 
NGC~1399 (whose nuclear star cluster is debatable) from Graham \& Spitler's
list. 
In Table~\ref{Tab1} we expand the above list of 10 (=12-2) galaxies for which black
holes and nuclear star clusters coexist.  We (i) provide masses for an
additional three galaxies (NGC~1300, NGC~2549 and NGC~3585, see
Figure~\ref{Fig1}) to give a total of 13, 
(ii) update the mass of the nuclear star cluster in NGC~4026, and (iii)
tabulate upper limits on the star cluster masses for a further seven galaxies.
Also provided in Table~\ref{Tab1} are the names of galaxies whose spheroid
mass is such that they are good candidates to house dual nuclei.

In passing, it is noted that the presence of nuclear star clusters with a
different stellar population and thus a different stellar $M/L$ ratio to the
surrounding bulge (e.g.\ Lotz et al.\ 2004; C\^ot\'e et al.\ 2006; Paudel et
al.\ 2011; den Brok et al.\ 2011, in prep.) may result in errors to the
derivation of the supermassive black hole mass if one is not careful.  We are
not, however, in a position to quantify this, and we take the quoted
supermassive black hole errors at face value.  As discussed in Graham \&
Spitler (2008), the uncertainty on the nuclear star
cluster masses is likely constrained to within a factor of $\sim$2.

% Rotationally flattened nuclear discs, with sizes much larger than nuclear star
% clusters (e.g., Balcells et al.\ 2007; Ledo et al.\ 2010), were not included
% by Graham \& Spitler (2009) nor are they included in Table~\ref{Tab1},
% although Seth et al.\ (2008b, see also Prieto et al.\ 2004) reveal that there
% can at times be a somewhat blurred distinction between these types of object.
% 
The nuclear star cluster masses and host galaxy velocity dispersions shown in
Ferrarese et al.\ (2006a), for 29 galaxies with $\sigma \la$120 km 
s$^{-1}$, have been included here to better populate the lower-mass end of our
($M_{\rm bh} + M_{\rm nc}$)-$\sigma$ diagram.  From that study, the four
nuclear star clusters with masses $\ga 10^8 M_{\odot}$ (VCC 1913,
1146, 1630, 1619) are reported to have half-light radii of 0.32, 0.50, 0.60
and 0.71 arcseconds, respectively (Ferrarese et al.\ 2006b). All of the
remaining nuclei sizes are less than $0.^{\prime\prime}25$, i.e.\ less than 20
pc adopting their Virgo cluster distance of 16.5 Mpc.  Ferrarese et al.\
(2006b) identified the first three of these four galaxies as hosting a small
scale nuclear disc, and they observed a very dusty nucleus in the lenticular
galaxy VCC~1619.  Through application of their S\'ersic-galaxy $+$
single-nucleus model, the flux which they assigned to their ``nuclear star
clusters'' is greater than that acquired when separating nuclear discs and
nuclear star clusters (e.g.\ Balcells et al.\ 2007).
% 
% Although we have not excluded these galaxies 
This explains the apparent
deviant nature of at least the first three of these four galaxies in
Figure~\ref{Fig2}b.

% The Sagittarius dwarf spheroidal galaxy (Sgr dSph) with $\sigma = 10\pm3$ km
% s$^{-1}$ is also included, 
% with the dense nuclear star cluster Sgr,N having $M_V = -7.8\pm0.2$ mag 
% and a mass of $6\times 10^6 M_{\odot}$
% (Bellazzini et al.\ 2008; Carretta et al.\ 2010),
% but see footnote 12 in Bellazzini and Fig.4 from Graham \& Spitler, which 
% gives $M/L_V = 1.5$ for a 6 Gyr old population with [Fe/H] = $-0.4$ 
% and thus a mass of $1.7\times 10^5 M_{\odot}$ using $M/L_V = 1.5$. 
% XXX??

\section{Results and Discussion}\label{Sec-R_D}

\begin{figure*}
\includegraphics[angle=270,scale=0.66]{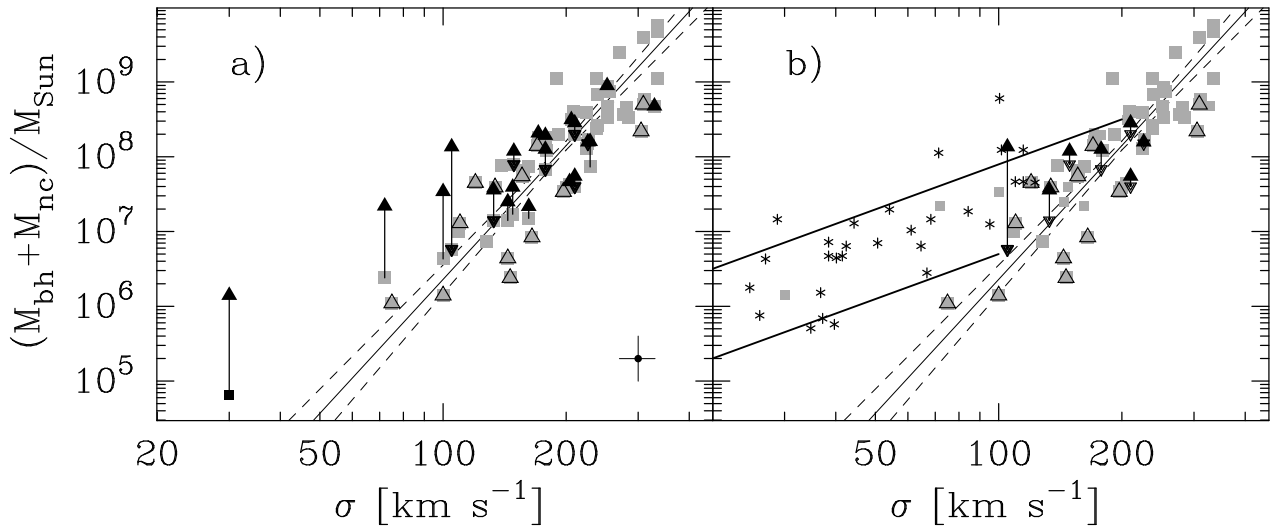}
\caption{
Panel a) 64 gray squares define the $M_{\rm bh}$--$\sigma$ relation from
Graham et al.\ (2011), shown by the thin line, 13 black arrows show how points
will move if the nuclear star cluster mass $M_{\rm nc}$ is added, while 
double-headed arrows are used for the 7 nuclear star clusters that only have an
upper limit to their mass. The 13 open triangles mark galaxies which may have
a nuclear star cluster that could move the points higher in the diagram (see Table~1).
A representative error bar is shown in the bottom right corner. 
Panel b) The 13 galaxies with known nuclear cluster masses are now shown by only
one single gray square.  
We have also now included, shown by the stars, 29 galaxies with nuclear cluster masses 
from Ferrarese et al.\ (2006a, their figure~2b). 
The two heavy black lines have a slope of 2. 
}
\label{Fig2}
\end{figure*}

Expanding upon the $(M_{\rm bh}+M_{\rm nc})$--$\sigma$ diagram from Graham et al.\ (2011,
their figure~8), especially at the
low-$\sigma$ end through the inclusion of the ($M_{\rm nc}, \sigma$)
data from Ferrarese et al.\ (2006a) proves to be rather revealing.  
Figure~\ref{Fig2} appears to display two markedly different 
slopes.  While the slope at the high-$\sigma$ end is around 5 for the ``core'' 
galaxies (Ferrarese \& Merritt 2000; Hu 2008; Graham et al.\ 2011), the slope at 
the low-$\sigma$ end is seen to be roughly consistent with a value of 2.  
Given that the efficiency of feedback from star clusters
and massive black holes is different, it is probably preferable
to separate their masses when considering slopes in $M$--$\sigma$ diagram. 
 
Fitting the ordinary least squares bisector regression {\sc SLOPES} (Feigelson
\& Babu 1992) --- a code which is not sensitive to measurement uncertainties
--- to the (13+29) nuclear stellar masses and associated velocity dispersions
mentioned in the previous section gives a slope of 2.14$\pm$0.31.
Although one may rightly wonder if this slope has been lowered by the
inclusion, at the high-$\sigma$ end, of nuclear star clusters which have been
partly eroded by massive black holes --- if the scenario proposed by Bekki \&
Graham (2010) is correct.  It is however the case that the four stellar nuclei with
masses $\ga 10^8 M_{\odot}$ do increase the measured slope.  Removing these four
objects results in a slope of 1.78$\pm$0.24 (and an intercept at 70 km s$^{-1}$
of 6.83$\pm$0.08), in remarkable agreement with the expected value of
1.74$\pm$0.52 (see section~\ref{Sec_light}) based on a smaller independent
data set. 
Using the bisector regression {\sc BCES} from Akritas \& Bershady (1996), 
and assuming a 10 and 50 per cent uncertainty on the velocity dispersion 
and the nuclear star cluster mass, respectively, gives a near identical 
slope and intercept of 1.73$\pm$0.23 and 6.83$\pm$0.07.  While varying the 
uncertainty on the velocity dispersion by a factor of 2 has almost no affect
on the fit, 
increasing the uncertainty on the nuclear star cluster mass to a factor of 2
yields the relation 
% a slope of 1.57$\pm$0.24 while leaving the intercept unchanged. 
\begin{equation}
\log \left[\frac{M_{\rm nc}}{M_{\odot}} \right] =
(1.57\pm0.24)\log \left[\frac{\sigma}{70\, {\rm km\, s}^{-1}}\right] + (6.83\pm0.07). 
\end{equation}

Figure~\ref{Fig2} suggests that nuclear star clusters do not cleary define 
an offset parallel relation that is disconnected from the distribution of black
holes in the $M$--$\sigma$ diagram, as suggested by Ferrarese et al.\ (2006a)
who had found that $M_{\rm nc}$--$\sigma^{4.27\pm0.61}$.  
Excluding what are likely to be nuclear stellar discs from four galaxies
studied by Ferrarese et al.\ (2006a; although see Prieto et al.\ 2004 and 
Seth 2008b), while including an additional 13 nuclear
star clusters in galaxies with velocity dispersions over a much larger
baseline, reaching out to $\sim$200 km 
s$^{-1}$, we have found a notably shallower $M{\rm nc}$--$\sigma$ relation.
The previous relation had inspired some to adapt the momentum-conserving
arguments of Fabian (1999; see also King \& Pounds 2003 and Murray et al.\
2005) which had been used to explain why an $M_{\rm bh}$--$\sigma^4$ relation
might arise. 
This nuclear cluster feedback mechanism involving stellar winds to produce an
$M_{\rm nc} \propto \sigma^4$ scaling relation may therefore require some
modification (McLaughlin et al.\ 2006; McQuillin \& McLaughlin 2012).
Relaxing the assumption of an isothermal sphere for the dark matter halo might
prove helpful. On the other hand, the results may be telling us that
(momentum) feedback is not relevant, which would be expected if the star
clusters were to have originated somewhere else and subsequently been
deposited into the spheroid, rather than coevolving there.

% XXX: Update the Graham \& Spitler (2009) figure 3, but draw a constant
% ratio above $\sim 10^{11} M_{\odot}$ for dry merging.

It is noted that the distribution of points defining the $(M_{\rm bh}+M_{\rm
nc})$--$\sigma$ relation seen in Figure~\ref{Fig2} may yet be shown to be
tracing an upper envelope at the low-$\sigma$ end.  For-example, non-nucleated
dwarf elliptical galaxies would reside below such an upper envelope if they do
not contain a supermassive black hole of sufficient mass (see also Batcheldor
2010 in regard to sample selection effects).

Finally, 
an argument can be made for expecting a slope (or upper envelope) 
at the low-$\sigma$ end of the $(M_{\rm bh}+M_{\rm nc})$--$\sigma$ diagram that 
is actually closer to 1 than 2.  While the data for galaxies with 
$M_{\rm bh} > 5\times10^7 M_{\odot}$ to $2\times10^8 M_{\odot}$ is 
roughly consistent with a constant $(M_{\rm bh}+M_{\rm nc})/M_{\rm sph}$ ratio 
(Marconi \& Hunt 2003; H\"aring \& Rix 2004),
Graham \& Spitler (2009, see their figure~3) found that the $(M_{\rm
  bh}+M_{\rm nc})/L$ ratio increases as one proceeds to lower
luminosities $L$ such that $L\propto (M_{\rm bh}+M_{\rm nc})^{5/3}$.
Subsequently, coupled with the relation $L\propto \sigma^2$, one has
that $(M_{\rm bh}+M_{\rm nc}) \propto \sigma^{6/5}$.  
Additional data plus a more detailed modelling of each galaxy's individual
stellar components, including inner and outer nuclear discs, 
will help to clarify this situation.

\subsection{Predictions for a bent $M_{\rm bh}$--$L$ and $M_{\rm bh}$--$M_{\rm
    sph}$ relation}\label{Sec_Predict}

We know that for massive elliptical galaxies $L \propto \sigma^5$ (Schechter
1980; Malumuth \& Kirshner 1981) and $M_{\rm bh} \propto \sigma^5$ (Merritt \&
Ferrarese 2001; Hu 2008; Graham et al.\ 2011).  Consistent with these
observations is the relation $M_{\rm bh} \propto L^{1.0}$ (Marconi \& Hunt
2003; Graham 2007) for galaxy samples dominated by massive elliptical
galaxies.  One may then ask what about the lower-mass galaxies (with $M_B \ga 
-20.5$ mag). As noted, these dwarf and intermediate-luminosity elliptical galaxies
have $L \propto \sigma^2$ (Davies et al.\ 1983; Matkovi\'c \& Guzm\'an 2005; de
Rijcke et al.\ 2005) while they also seem to follow the relation $M_{\rm bh}
\propto \sigma^5$ (Ferrarese \& Merritt 2000; Graham et al.\
2011).\footnote{The offset nature of barred / pseudobulge galaxies in the 
$M_{\rm bh}$--$\sigma$ diagram (Graham 2008a; Hu 2008) appears to be an
unrelated phenomenon.}  Consequently, one should find that $M_{\rm bh} \propto
L^{2.5}$ for elliptical galaxies with $M_B \ga -20.5$ mag ($M_{\rm bh} \la 
5 \times 10^7$ -- $2\times10^8 M_{\odot}$).  That is, the $M_{\rm bh}$--$L$
relation may be broken or curved, and the $M_{\rm bh}/L$ and $M_{\rm
bh}/M_{\rm sph}$ ratios may not be approximately constant values at these
lower masses.  
This has nothing to do with pseudobulges nor the alleged divide between
elliptical and dwarf elliptical galaxies at $M_B = -18$ mag (see the review in
Graham 2012a). 
Further support for the above suggestion stems from the observation 
that the luminosity-(S\'ersic index) relation is linear (e.g.\ Graham \&
Guzm\'an 2003) while the $M_{\rm bh}$--(S\'ersic index) relation is curved or
broken (Graham \& Driver 2007).  Consistency would require that the $M_{\rm
bh}$-luminosity relation be broken too. 

Spheroids fainter than $M_B = -20.5$ mag are the dominant spheroid population in the
universe, and it is claimed here that past work on the
$M_{\rm bh}$--$M_{\rm sph}$ and $M_{\rm bh}$--L relations have 
been severely biased by the sample selection of luminous spheroids likely
built in `dry' merger events. As such, the current near-linear 
$M_{\rm bh}$--$M_{\rm sph}$ and $M_{\rm bh}$--L relations 
(e.g.\ Marconi \& Hunt 2003; H\"aring \& Rix 2004; Graham 2007) should 
not be used to constrain the growth mechanism of supermassive black holes in
galaxies (beyond simple addition in `dry' merger events).
This prediction, with significant implications for galaxy formation if true, 
will be investigated further in (Graham 2012b).

\section{acknowledgment}

The author wishes to have it acknowledged that more than five months elapsed
between submitting his manuscript and receiving a referee letter. 
This research was supported by Australian Research Council
grants DP110103509 and FT110100263. 
Graham thanks the organisers of the conference ``Central
Massive Objects: The Stellar Nuclei-Black Hole Connection'', June
22-25, 2010, ESO Headquarters, Garching, Germany, where this work was
first presented. 
% 
% This research has made use of NASA’s Astrophysics Data System (ADS)
% Bibliographic Services, the NASA/IPAC Extragalactic Database (NED),
% and the HyperLeda database (http://leda.univ-lyon1.fr).

\label{lastpage}
\end{document}